\documentclass[preprint,showpacs,preprintnumbers,amsmath,amssymb]{revtex4-2}
\usepackage{bm}
\usepackage{amsmath}
\usepackage{amssymb}
\usepackage{amsfonts}
\usepackage{graphicx}
\begin{document}

\title{``To renormalize or not to renormalize ?'' 
in the proton-deuteron scattering calculations}

\author{H.~Wita{\l}a}
\affiliation{M. Smoluchowski Institute of Physics, 
Faculty of Physics, Astronomy and Applied Computer Science,
Jagiellonian University, PL-30059 Krak\'ow, Poland}

\author{J.~Golak}
\affiliation{M. Smoluchowski Institute of Physics, 
Faculty of Physics, Astronomy and Applied Computer Science,
Jagiellonian University, PL-30059 Krak\'ow, Poland}

\author{R.~Skibi\'nski}
\affiliation{M. Smoluchowski Institute of Physics, 
Faculty of Physics, Astronomy and Applied Computer Science,
Jagiellonian University, PL-30059 Krak\'ow, Poland}

\date{\today}

\begin{abstract}
  We discuss  two  approaches which, by applying the screening method, permit one
  to include the long range
  proton-proton (pp) Coulomb force
  in  proton-deuteron (pd) momentum-space scattering
  calculations. 
  In the first one,  based on
  Alt-Grassberger-Sandhas (AGS)  equation, presented  
   in Phys. Rev. C{\bf{71}}, 054005 (2005) and   
   {\bf{73}}, 057001 (2006), one needs to renormalize
   elastic scattering amplitude before calculating observables.
   In the second treatment, proposed by us in
   Eur. Phys. Journal A {\bf{41}}, 369 (2009), {\bf{41}}, 385 (2009), and
   arXiv:2310.03433 [nucl.th], 
   this renormalization is avoided. For the proton induced deuteron breakup reaction
   both approaches require
   renormalization of the corresponding transition amplitudes.
   We derive the basic equations underlying both methods under the assumption that
   all contributing  partial wave states are included and explain 
   why in our approach renormalization of the elastic scattering amplitude is
   superfluous.  
   We show that in order to take into account in the
   screening limit all partial waves it is required that four
   additional terms, based on
	the 3-dimensional and partial-wave projected pp Coulomb 
   t-matrices,  identical  for  
    both approaches, must appear in transition amplitudes. 
   We investigate importance of these terms for elastic pd scattering below
   the breakup threshold.
\end{abstract}


\maketitle \setcounter{page}{1}

The Hamlet-like question in the title arose when two preprints
~\cite{wita_preprint} and ~\cite{delt_preprint}, both dealing with the  
problem how to include the long range proton-proton (pp)
Coulomb force in momentum space pd scattering calculations through 
a screened Coulomb interaction, were posted.
 The arguments presented in
~\cite{delt_preprint} show that in the well established  approach of  
 Refs.~\cite{delt2005el,delt2005br} the interplay of the pp Coulomb
potential and the deuteron bound state pole  in the neutron-proton t-matrix 
makes renormalization of the elastic scattering transition
amplitude necessary prior to calculating observables.
Contrary to that, in our approach presented in
~\cite{wita_preprint,elascoul,brcoul}, one  
avoids  such renormalization. In the following we explain 
similarities and differences of both treatments and  provide justification why
the renormalization in our method for elastic scattering  is unnecessary.
 We also discuss a very
 important problem, indispensable in any treatment of the long-range Coulomb force: 
 how to take into account, in addition to partial waves
utilised when solving corresponding three-nucleon (3N) scattering equations, 
all higher partial wave states.

Let us start with the well established approach of
 Refs.~\cite{delt2005el,delt2005br}
based on the AGS equation for the pd transition operator
$U$ ~\cite{AGS,gloeckle83}: 
\begin{eqnarray}
 U\left| {\Phi } \right\rangle  = PG_0^{-1}\left| {\Phi } \right\rangle 
 + PtG_0U \left| {\Phi } \right\rangle ~,
 \label{eq1}
\end{eqnarray}
 where $ P $ is defined in terms of transposition operators,
 $ P =P_{12} P_{23} + P_{13} P_{23} $, $G_0$ is the free 3N propagator
 and $ |\Phi>$ is the initial state composed of a deuteron and a momentum
 eigenstate of the proton. 
 The t-matrix $t$ is a solution of the 2-body Lippmann-Schwinger (LS) equation, 
 with the interaction which contains in case of the pp system in
 addition to the nuclear part also
 the Coulomb pp force (assumed to be screened and parametrized
 by some parameter $R$).  
 If the state $U| \Phi>$ is known, the elastic pd scattering amplitude
 $< \Phi~'|U| \Phi>$, with $|\Phi~'>$ being the final pd state,  can be obtained
 by quadratures in the standard manner.

 In our approach we use the breakup operator T defined as:
\begin{eqnarray}
  T= tG_0U ~.
 \label{eq2}
\end{eqnarray} 
It fulfills the 3N Faddeev equation which,  
when nucleons interact
with pairwise forces only, is given by ~\cite{physrep96,gloeckle83}:
\begin{eqnarray} T| \Phi> =  t P | \Phi> +  t P G_0 T|  \Phi> ~.
  \label{eq3}
\end{eqnarray}
The above form of the Faddeev equation ensures that the T operator
reflects directly the properties of the t-matrix. Here the elastic scattering
amplitude is calculated from solutions of
(\ref{eq3}) by ~\cite{gloeckle83,physrep96}:
\begin{eqnarray}
 \left\langle {\Phi '} \right|U\left| {\Phi } \right\rangle
 &=&
 \left\langle {\Phi '} \right|PG_0^{ - 1} \left| {\Phi }
 \right\rangle   + \left\langle {\Phi '} \right| PT\left| {\Phi }
 \right\rangle ~,
 \label{eq4}
\end{eqnarray}
and the transition amplitude for 
 breakup $<\Phi_0|U_0|\Phi>$ is expressed 
in terms of $T\left| {\Phi } \right\rangle$ by~\cite{gloeckle83,physrep96}
\begin{eqnarray}
 \left\langle {\Phi _0 } \right|U_0 \left| {\Phi } \right\rangle  &=&
 \left\langle {\Phi _0 } \right|(1 + P)T\left| {\Phi } \right\rangle ~,
\label{eq5a}
 \end{eqnarray}
where $ | \Phi_0> = | \vec p \,
\vec q \, m_1 m_2 m_3 \nu_1 \nu_2 \nu_3 > $ is the  state of three free
outgoing nucleons. In the approach based on the AGS equation the transition
amplitude for breakup is given also by Eq. (\ref{eq5a}) but with T replaced by U.

The AGS, (\ref{eq1}), as well as the Faddeev, (\ref{eq3}),
    equations are solved in the momentum-space partial-wave basis $|pq \bar{\alpha}>$:
\begin{eqnarray}
|p q \bar{\alpha}> \equiv  |pq(ls)j(\lambda
\frac {1} {2})I (jI)J (t \frac {1} {2})T> ~,
\label{eq5}
\end{eqnarray}
where one can differentiate between the partial wave states $|pq\alpha>$ with
total 2N angular momentum $j$ below some value $j_{max}$: $j \le
j_{max}$,
 in which the nuclear, $V_N$, as well as the pp screened Coulomb
interaction, $V_c^R$
 (in isospin $t=1$ states only), act,  and the
states $|pq\beta>$ with $j > j_{max}$, for which only the screened Coulomb
force $V_c^R$ is present  
in the pp subsystem. Incorporation  of the $|pq\beta>$ states is indispensable
due to the long range nature of the pp Coulomb force and the necessity to
perform finally the screening limit $R \to \infty$. 
 In the following we derive  for both approaches the equations
in a subspace restricted to $|pq\alpha>$ states only, which, however, incorporate all
contributions from the complementary subspace of $|pq\beta>$ states. 
 The states $|pq\alpha>$ and
 $|pq\beta>$ form together a complete system of states
 (in the following we use shorthand notation
$ \sum\limits_{\alpha}
 {\int {p^2 dpq^2 dq\left| {pq\alpha} \right\rangle \left\langle
     {pq\alpha} \right|} }\equiv   {\left|
     {\alpha} \right\rangle \left\langle {\alpha} \right|}  $):
\begin{equation}
\int {p^2 dpq^2 dq} (\sum\limits_\alpha  {\left| {pq\alpha }
\right\rangle } \left\langle {pq\alpha } \right| +
\sum\limits_\beta  {\left| {pq\beta } \right\rangle } \left\langle
           {pq\beta } \right|) =
   {\left| {\alpha} \right\rangle \left\langle {\alpha} \right|} +
   {\left| {\beta} \right\rangle \left\langle {\beta} \right|} 
    = {\rm I} \\ ~
\label{eq6}
\end{equation}
where I is the identity operator.

Let us start with our approach. 
Projecting Eq.~(\ref{eq3}) for $ T| \Phi >$ on the $|pq\alpha>$ and
$|pq\beta>$ states one gets the following system of coupled
integral equations ~\cite{wita_preprint}: 
\begin{eqnarray}
 \left\langle {pq\alpha } \right|T\left| {\Phi } \right\rangle  &=&
 \left\langle {pq\alpha } \right|t_{N + c}^R P\left| {\Phi } \right\rangle 
 + \left\langle {pq\alpha } \right|t_{N + c}^R PG_0 
 {\left| {\alpha '} \right\rangle \left\langle
 {\alpha '} \right|}  T\left| {\Phi } \right\rangle \cr
 &+& \left\langle {pq\alpha } \right|t_{N + c}^R PG_0 
  {\left| {\beta '} \right\rangle \left\langle
     {\beta '} \right|}  T\left| {\Phi } \right\rangle
 \label{eq7} ~, \\
 \left\langle {pq\beta } \right|T\left| {\Phi } \right\rangle  &=&
 \left\langle {pq\beta } \right|t_c^R P\left| {\Phi } \right\rangle 
  + \left\langle {pq\beta } \right|t_c^R PG_0 
  \left| {\alpha '} \right\rangle
  \left\langle {\alpha '} \right|  T\left| {\Phi } \right\rangle  ~,
 \label{eq8}
\end{eqnarray}
where $t_{N+c}^R$ and $t_c^R$ are t-matrices generated by
the interactions $V_N+V_c^R$ and $V_c^R$, respectively.

Inserting $<pq\beta|T|\Phi>$ from (\ref{eq8}) into (\ref{eq7}) and using
(\ref{eq6}) one gets:
\begin{eqnarray}
 \left\langle {pq\alpha } \right|T\left| {\Phi } \right\rangle  &=&
 \left\langle {pq\alpha } \right|t_{N + c}^R P\left| {\Phi } \right\rangle
 + \left\langle {pq\alpha } \right|t_{N + c}^R PG_0 t_c^{R 3d} P\left| {\Phi }
\right\rangle \cr
  &-& \left\langle {pq\alpha } \right|t_{N + c}^R PG_0 
  \left| {\alpha '} \right\rangle
  \left\langle {\alpha '} \right| t_c^R P\left| {\Phi } \right\rangle
\cr
&+& \left\langle {pq\alpha } \right|t_{N + c}^R PG_0
\left| {\alpha '} \right\rangle
  \left\langle {\alpha '} \right| T\left| {\Phi }
  \right\rangle \cr
  &+& \left\langle {pq\alpha } \right|t_{N + c}^R PG_0 t_c^{R 3d} PG_0
\left| {\alpha '} \right\rangle
  \left\langle {\alpha '} \right| T\left| {\Phi } \right\rangle \cr
  &-& \left\langle {pq\alpha } \right|t_{N + c}^R PG_0
               \left| {\alpha '} \right\rangle
  \left\langle {\alpha '} \right| t_c^R PG_0 
   \left| {\alpha'' } \right\rangle
  \left\langle {\alpha'' } \right| T\left| {\Phi } \right\rangle ~.
  \label{eq9}
 \end{eqnarray}
This is a set of coupled integral equations in the space of the 
$\left| {{\alpha } } \right\rangle$ states, which exactly
incorporates the contributions of
the pp Coulomb interaction from all partial wave states up to
infinity. It is clear that there is a price to pay for taking into account all
states $|pq\beta>$: the necessity to work with the 3-dimensional
Coulomb t-matrix $t_c^{R 3d}$, obtained by solving the  3-dimensional
LS equation ~\cite{skib2009}. 

Presently it is practically impossible to solve Eq.~(\ref{eq9}) in its full 
glory. The reason are drastic amount of computer resources and of computer time
required to calculate the second and the fifth terms with the 3-dimensional
Coulomb t-matrix. Luckily enough, one can rather easily eliminate them at the
expense of increasing the basis of $\left| {{\alpha } } \right\rangle$
states. Namely, extending the set $\left| {{\alpha } } \right\rangle$ 
 by adding  channels with higher angular
 momenta, in which only the pp Coulomb interaction is present, permits one to
 completely neglect the four terms in (\ref{eq9}) due to their mutual
 cancellation: the second with the third and the fifth with the sixth term.
 The set (\ref{eq9}) is then reduced to: 
\begin{eqnarray}
 \left\langle {pq\alpha } \right|T\left| {\Phi } \right\rangle  &=&
 \left\langle {pq\alpha } \right|t_{N + c}^R P\left| {\Phi } \right\rangle
 + \left\langle {pq\alpha } \right|t_{N + c}^R PG_0 
 \left| {\alpha '} \right\rangle \left\langle
 {\alpha '} \right| T\left| {\Phi } \right\rangle  ~ ,
 \label{eq10} 
\end{eqnarray}
which is a basic equation in our approach (in ~\cite{wita_preprint}
called a simplified one).
It has identical structure as so frequently used 3N Faddeev equation
for neutron-deuteron (nd) scattering~\cite{physrep96}.

To calculate in our approach the elastic scattering transition
amplitude  one needs  in (\ref{eq4}) the second term 
 $\left\langle {\vec p\vec q~} \right|T\left| {\Phi }
\right\rangle$ composed of low ($\alpha$) and high ($\beta$)  partial wave 
contributions for $ T | \Phi>$. Using the completeness relation
(\ref{eq6}) one gets:
\begin{eqnarray}
\left\langle {\vec p\vec q~} \right| T \left| {\Phi }
\right\rangle  &=& \left\langle {\vec p\vec q~} \right| 
        {\alpha '} \left\rangle \right\langle {\alpha '} \left|
        T \right| {\Phi } \left\rangle   
 + \left\langle {\vec p\vec q~}
\right| t_c^{R 3d} P \left| {\Phi } \right\rangle
- \left\langle {\vec
  p\vec q~} \right| {\alpha '}
    \right\rangle \left\langle {\alpha '}
    \right| t_c^R P\left| {\Phi } \right\rangle \cr 
  &+& \left\langle
            {\vec p\vec q~} \right|t_c^{R 3d} PG_0 \left| {\alpha '}
            \right\rangle \left\langle
{\alpha '} \right| T\left| {\Phi } \right\rangle 
- \left\langle {\vec p \vec q~} \right|
  {\alpha '} \left\rangle \right\langle
{\alpha '} \left| t_c^R PG_0  \right| {\alpha ''}
\left\rangle \right\langle {\alpha ''} \left| T\right|
{\Phi } \left\rangle  \right .  ~.
\label{eq11}
\end{eqnarray}
To account correctly for contributions from 
$\left| {{\beta } } \right\rangle$ states again 
four terms are required, two of which contain  3-dimensional
Coulomb t-matrix. The first one, $\left\langle {\vec p\vec q~}
\right| t_c^{R 3d} P \left| {\Phi } \right\rangle$, corresponds to 
the amplitude of the Rutherford point-deuteron pd scattering  and the second one,
 $\left\langle
            {\vec p\vec q~} \right|t_c^{R 3d} PG_0 \left| {\alpha '}
            \right\rangle \left\langle
                          {\alpha '} \right| T\left| {\Phi } \right\rangle$,
 is a modification of the first one by nucleon-nucleon (NN) interactions.

Now we derive analogous relations in the approach based on the
AGS equation.
Projecting (\ref{eq1})  on the $|pq\alpha>$ and
$|pq\beta>$ states 
 and  using shorthand notation: 
$$ \sum\limits_{\alpha,\tilde{\alpha}}
{\int {p^2 dp q^2 dq p'^2 dp' \left| {pq\alpha} \right\rangle
    t^{\alpha \tilde{ \alpha}}(p,p';E-\frac {3} {4m} q^2)
 G_0 \left\langle {pq \tilde {\alpha}} \right|}}
 \equiv   \left|
   {\alpha} \right\rangle t^{\alpha} G_0 \left \langle {\alpha} \right|$$
 one gets the following system of coupled integral equations:
\begin{eqnarray}
 \left\langle {pq\alpha } \right|U\left| {\Phi } \right\rangle  &=&
 \left\langle {pq\alpha } \right| PG_0^{-1}\left| {\Phi } \right\rangle 
 + \left\langle {pq\alpha } \right| P \left| {\alpha ' } \right\rangle 
 t_{N + c}^{R { \alpha '}} G_0 
   {\left\langle
 {\alpha '} \right|}  U\left| {\Phi } \right\rangle \cr
  &+& \left\langle {pq\alpha } \right| P \left| {\beta ' } \right\rangle 
  t_{c}^{R  { \beta '}} G_0 
  \left\langle
     {\beta '} \right|  U\left| {\Phi } \right\rangle
 \label{eq12} ~, \\
 \left\langle {pq\beta } \right|U\left| {\Phi } \right\rangle  &=&
   \left\langle {pq\beta } \right| PG_0^{-1}\left| {\Phi } \right\rangle    
 + \left\langle {pq\beta } \right| P \left| {\alpha ' } \right\rangle 
 t_{N+c}^{R {\alpha ' }} G_0 
 \left\langle {\alpha '} \right|  U\left| {\Phi } \right\rangle \cr
&+& \left\langle {pq\beta } \right| P \left| {\beta ' } \right\rangle 
 t_{c}^{R {\beta' }}  
 \left\langle {\beta '} \right|  P\left| {\Phi } \right\rangle \cr
 &+& \left\langle {pq\beta } \right| P \left| { \beta ' } \right\rangle 
 t_{c}^{R {\beta' }} G_0
  \left\langle { \beta ' } \right| P \left| {\alpha ' } \right\rangle 
 t_{N+c}^{R {\alpha ' }} G_0 
 \left\langle {\alpha '} \right|  U\left| {\Phi } \right\rangle
 ~.
 \label{eq13}
\end{eqnarray}

Inserting $<pq\beta|U|\Phi>$ from (\ref{eq13}) into (\ref{eq12}) and using
(\ref{eq6}) one gets finally:
\begin{eqnarray}
  \left\langle {pq\alpha } \right|U\left| {\Phi } \right\rangle  &=&
\left\langle {pq\alpha } \right| PG_0^{-1}\left| {\Phi } \right\rangle   
+ \left\langle {pq\alpha } \right| P \left| {\alpha ' } \right\rangle 
 t_{N + c}^{R { \alpha '}} G_0 
   {\left\langle
 {\alpha '} \right|}  U\left| {\Phi } \right\rangle  \cr
   &-& \left\langle {pq\alpha }  \right| P \left| {\alpha ' } \right\rangle 
   t_{c}^{R {\alpha ' }} 
  \left\langle {\alpha '} \right| P\left| {\Phi } \right\rangle
  + \left\langle {pq\alpha }\right| P t_{c}^{R3d} P \left| {\Phi }
  \right\rangle \cr
  &-& \left\langle {pq\alpha } \right|P \left| {\alpha '} \right\rangle 
  t_{c}^{R { \alpha '}} G_0 {\left\langle {\alpha '} \right|} P
  \left| {\alpha '' } \right\rangle   t_{N + c}^{R {\alpha ''}} G_0 
  {  \left\langle {\alpha ''} \right|}  U\left| {\Phi } \right\rangle \cr
  &+&
\left\langle {pq\alpha } \right|P t_{c}^{R 3d} G_0 P
  \left| {\alpha '' } \right\rangle   t_{N + c}^{R {\alpha ''}} G_0 
  {  \left\langle {\alpha ''} \right|}  U\left| {\Phi } \right\rangle 
  ~.
  \label{eq14}
 \end{eqnarray}
This is a set of coupled integral equations in the space spanned by the 
$\left| {{\alpha } } \right\rangle$ states, analogous to (\ref{eq11})
in our approach.

Again, extending the set $\left| {{\alpha } } \right\rangle$ 
 by adding a finite number of channels with higher angular
 momenta,  leads to cancellations between  last four terms and 
 set (\ref{eq14}) is reduced to the following basic equation
 for approach based on AGS equation ~\cite{delt2005el,delt2005br}:
\begin{eqnarray}
 \left\langle {pq\alpha } \right| U \left| {\Phi } \right\rangle  &=&
 \left\langle {pq\alpha } \right| P G_0^{-1} \left| {\Phi } \right\rangle
 + \left\langle {pq\alpha } \right| P \left| {\alpha ' } \right\rangle 
 t_{N + c}^{R { \alpha '}} G_0 \left\langle {\alpha '} \right|
 U \left| {\Phi } \right\rangle  ~ .
 \label{eq15} 
\end{eqnarray}

To calculate the elastic scattering transition
amplitude $\left\langle {\Phi '} \right|U\left| {\Phi } \right\rangle$
one needs $\left\langle {\vec p\vec q~} \right|U\left| {\Phi }
\right\rangle$ composed of low ($\alpha$) and high ($\beta$)  partial wave 
contributions for $ U | \Phi>$. Employing the completeness relation
(\ref{eq6}) and Eq.~(\ref{eq14}) one gets:
\begin{eqnarray}
\left\langle {\vec p\vec q~} \right|U\left| {\Phi }
  \right\rangle  &=& \left\langle {\vec p\vec q~} \right|
 PG_0^{-1}  \left| {\Phi }  \right\rangle 
+ \left\langle {\vec p\vec q~} \right| P \left| {\alpha ' } \right\rangle 
 t_{N + c}^{R { \alpha '}} G_0 {\left\langle
   {\alpha '} \right|}  U\left| {\Phi } \right\rangle
 \cr 
 &-& \left\langle {\vec p\vec q~} \right|P \left| {\alpha ' } \right\rangle 
 t_{c}^{R { \alpha '}}  
   {\left\langle {\alpha '} \right|} P \left| {\Phi } \right\rangle  
   + \left\langle {\vec p\vec q~} \right|
   P t_{c}^{R 3d} P \left| {\Phi } \right\rangle
 \cr 
 &-& \left\langle {\vec p\vec q~} \right| P \left| {\alpha '} \right\rangle 
  t_{c}^{R { \alpha '}} G_0 {\left\langle {\alpha '} \right|} P
  \left| {\alpha '' } \right\rangle   t_{N + c}^{R {\alpha ''}} G_0 
  {  \left\langle {\alpha ''} \right|}  U\left| {\Phi } \right\rangle 
  \cr  
&+& \left\langle {\vec p\vec q~} \right|   P t_{c}^{R 3d} G_0 P
\left| {\alpha '' } \right\rangle 
 t_{N+c}^{R { \alpha ''}} G_0   
 {\left\langle {\alpha ''} \right|} U \left| {\Phi } \right\rangle
  ~.
\label{eq16}
\end{eqnarray}
Using relation (\ref{eq2}) between $U$ and $T$ one finds
that indeed amplitudes and thus also observables are the same
in both treatments.

It should be emphasized that only by extending the set of
$\left| {{\alpha } } \right\rangle$ states is it possible to neglect
in (\ref{eq9}) and (\ref{eq14}) the
 terms which contain the 3-dimensional Coulomb t-matrices,
 and to reduce the problem in both
 approaches to numerically well treatable
 equations (\ref{eq10}) and (\ref{eq15}).
 The indication that cancellations takes place
is given by convergence of predictions with respect to the
total angular momentum in the two-nucleon (2N) subsystem $j_{max}$, 
which defines
the set of $\left| {{\alpha } } \right\rangle$ states. It will be denoted in
the following by $j_s\, j_{max}$ with $j_s$ being the largest angular momentum in which
 the 2N interaction acts~\cite{wita_preprint}.

It is  evident that a correct treatment of the Coulomb force 
in both approaches requires inclusion of four additional terms in the
elastic (and also breakup) transition amplitudes (the last four terms
in (\ref{eq11}) and (\ref{eq16})).

\begin{figure}
\includegraphics[scale=0.8]{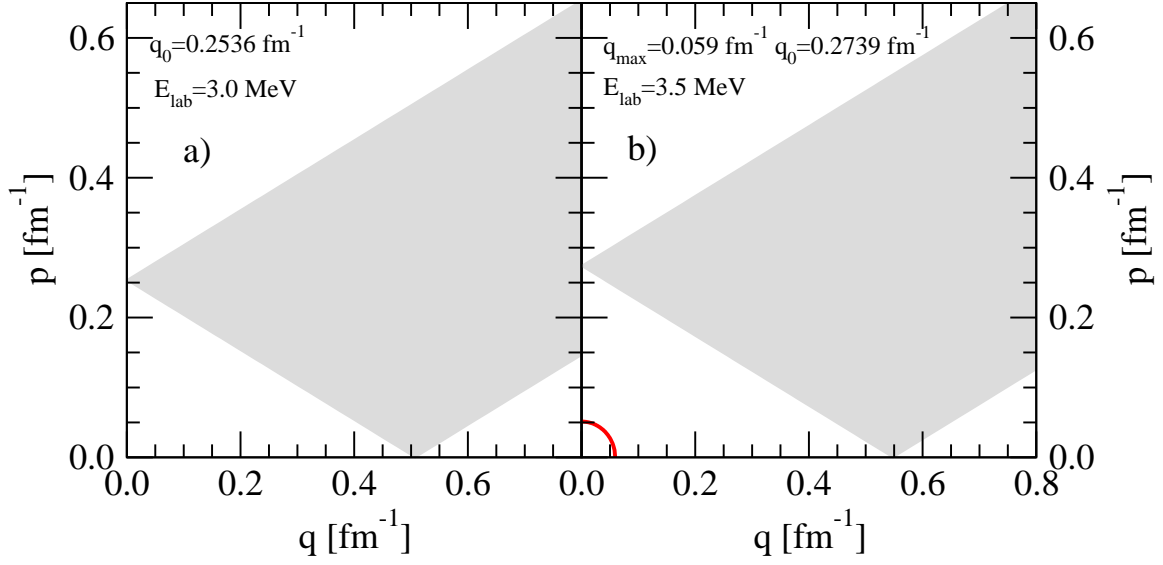}
  \caption{(color online) Regions of the Jacobi momenta
    $q$ and $p$ values  in
    $(q-p)$ plane which contribute to the
  breakup reaction ((red) solid line at $E=3.5$~MeV, showing ellipse
  $\frac {p^2} {m} + \frac {3} {4m} q^2 = \frac{3} {4m} q_{max}^2
  = \frac{3} {4m} q_0^2 + E_d $ ) and elastic scattering
    ($<\Phi' |P T | \Phi>$  term) (gray highlighted region)
    at the incoming nucleon laboratory energy $E=3.0$ and $3.5$~MeV.
 }
 \label{fig1}
\end{figure}

It was shown in ~\cite{delt_preprint} (see also
 ~\cite{delt2005el,delt2005br} and references therein)
that in the treatment based on AGS equation (\ref{eq15}) the
elastic scattering transition amplitude acquires in the screening limit
$R \to \infty$ an infinitely oscillating phase factor 
and must be renormalized before calculating observables. 
As a consequence, each term in (\ref{eq16}) containing
 $U\left| {\Phi } \right\rangle$ has to be renormalized. 
 In our approach we solve instead of AGS  
  the 3N Faddeev equation (\ref{eq10}) for the 
 $\left\langle {p q \alpha} \right| PT\left| {\Phi } \right\rangle$ states,
 from which later elastic scattering transition amplitude is calculated.
 In this way we avoid the main source 
 of the oscillating phase factor described in ~\cite{delt_preprint}
 and the necessity of renormalization of the elastic
 scattering amplitude.   
Additionally, the structure of 3N Faddeev equation guarantees
  that their solutions inherit 
 properties from the two-nucleon t-matrices providing thus 
  an additional argument that renormalization is
 redundant. 
 Namely, the properties of t-matrices generated by the screened Coulomb force
 alone 
 (in the case of partial wave decomposed t-matrices also those generated  by a
 combination of Coulomb and nuclear parts) 
 as well as their
 screening limits were studied theoretically  in the past in numerous papers  
 ~\cite{Alt78,chen72,kok1980,ford1964,ford1966,taylor1,taylor2,kok1981,haer1985}
 and later some of these properties  were confirmed numerically
 in ~\cite{skib2009}. The most important finding was that such off-shell
  t-matrices have a well defined screening limit while
 the half- and on-shell ones acquire in this limit
 an infinitely oscillating phase factor. 
At the same time, the elastic pd scattering amplitude
  gets contributions of  
  $\left\langle {pq\alpha } \right|T\left| {\Phi } \right\rangle$ states
  only from the off-shell region of the Jacobi momenta magnitudes $q$ and $p$
   in ($q-p$) plane: 
  $\frac {p^2} {m} + \frac {3} {4m} q^2 \ne \frac{3} {4m} q_{max}^2 = 
 \frac{3} {4m} q_0^2 +E_d$, 
where $m$ is the nucleon mass,
$E_d$ is the (negative) deuteron binding energy, 
and $q_0$ is the magnitude of the relative pd momentum. 
That off-shell
 region of $q-p$ values does not
 overlap with the ellipse from which half-on-shell
 contributions to the breakup reaction come. 
 In Fig. \ref{fig1} we exemplify that off-shell part 
  and the separation of the
 breakup and elastic scattering regions in the $(q-p)$ plane
 for the energy of a pd system $E=3.5$~MeV, which is slightly above
 the breakup threshold and for which both reactions are possible, and at
 $E=3.0$~MeV, which is below the
 breakup threshold and for which only elastic scattering is allowed.
 The fact that elastic pd scattering 
requires only off-shell solutions of the Faddeev equations and that 
the off-shell two-nucleon t-matrices have a well defined screening limit 
is the reason why in our method no renormalization of elastic scattering
amplitudes is needed. Contrary to that, the breakup amplitudes acquire
 the oscillating phase factor originating from half-shell t-matrices. 

\begin{figure}
  \includegraphics[scale=0.5]{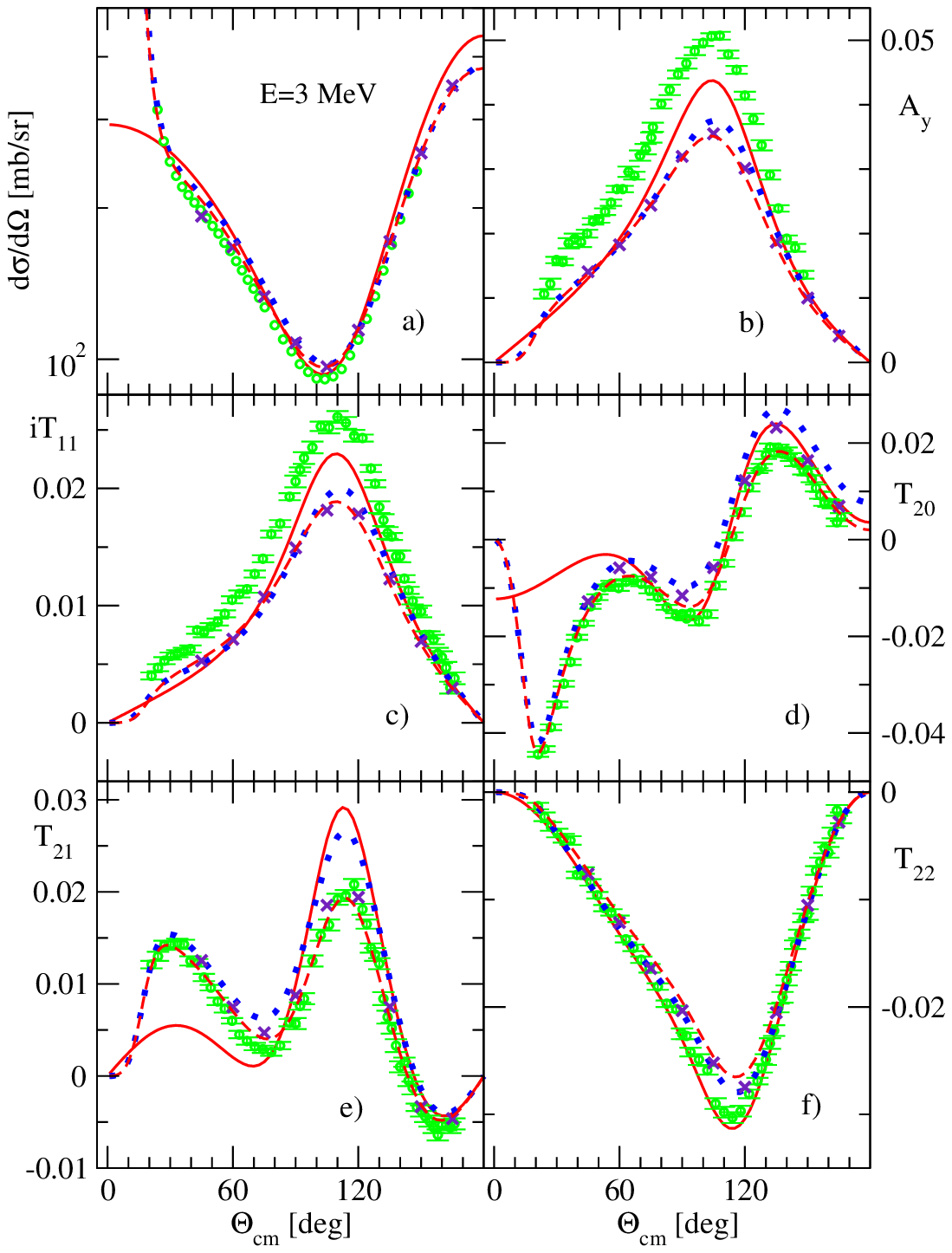}
\caption{(color online) Comparison of  data and predictions for the
pd scattering cross section $\frac {d\sigma} {d\Omega}$, 
proton vector $A_y$, deuteron vector $iT_{11}$ 
and deuteron tensor $T_{20}$, $T_{21}$, $T_{22}$ analyzing powers. 
	They are shown
as  functions of a c.m. proton scattering 
	angle $\Theta_{cm}$ and were calculated
 at the incoming proton laboratory energy $E=3.0$~MeV 
 with the approach based on Faddeev equation 
  (\ref{eq10}) and 
transition amplitude (\ref{eq11}). The exponentialy screened Coulomb
force ($R=40$~fm, $n=4$) and the AV18 potential~\cite{av18}
restricted to the $j \le 3$ partial waves have been applied.
 To solve Faddeev equation the set $j_s3j7$ of
 $ \left| {\alpha } \right\rangle$ states was used. 
 The red short dashed lines show the results when only the first 
 three terms in (\ref{eq11}) are taken into account.   
   The blue dotted lines are predictions  when also the fifth
  term in (\ref{eq11}) ($ - \left\langle {\vec p \vec q~} \right|
  {\alpha '} \left\rangle \right\langle
{\alpha '} \left| t_c^R PG_0  \right| {\alpha ''}
\left\rangle \right\langle {\alpha ''} \left| T \right|
 {\Phi } \left\rangle  \right.$) is included. 
  The pure Coulomb term
 $\left\langle {\Phi' } | Pt_cP| {\Phi } \right\rangle$
  was determined using the screening limit expresion for the off-shell
  3-dimensional Coulomb t-matrix (Eq.~(19) in Ref.~\cite{wita_preprint}).
 The indigo crosses show the results with all terms in (\ref{eq11}) included.
  The red solid lines are predictions for nd elastic scattering and 
 green circles represent the pd data from Ref.~\cite{shimizu1995}. 
 }
 \label{fig2}
\end{figure}

In order to compare results of two approaches and check that indeed
our method does not need the renormalization, 
 we applied our approach at a low proton energy below the breakup threshold,
 where effects of the pp Coulomb
force as well as contributions of different terms to the
elastic scattering amplitude are expected to be dominant and 
where also results of the AGS approach are available
at $E=3.0$~MeV ~\cite{delt2005el}. 
 In Fig.~\ref{fig2}  we show our
 predictions obtained with the AV18 NN potential~\cite{av18} and $j_s3j7$
 $ \left| {\alpha } \right\rangle $ basis 
at  $3.0$~MeV compared to existing
elastic scattering data for the  cross section and  analyzing powers.
The red short dashed line show results obtained with only the
first three terms in elastic scattering transition amplitude (\ref{eq11}),
which is the approximation used also in Ref.~\cite{delt2005el}.  
The red solid lines are predictions for neutron-deuteron scattering.
It is clear that in this region of energies  
the Coulomb force effects indeed
are large and dominant at all angles as evidenced by
comparing the red solid and short dashed lines.
 It is astonishing how good the overall 
description of tensor analyzing power data is in spite of their small magnitudes
of $\approx 1 \%$. The vector
analyzing powers $A_y$ and $iT_{11}$ are underestimated by theory what is
very well known in the literature under the
name ``low energy analyzing power puzzle''.
Even more interesting is the good agreement for practically
all shown observables, with the exception of $A_y$ and $iT_{11}$,
 between our $3.0$~MeV 
results and the predictions based on the AGS approach, 
as far as it can be judged from
Fig.~9 of  Ref.~\cite{delt2005el}. This good agreement strongly supports
the statement that both approaches have to provide the same predictions 
for all observables and that in our approach renormalization of the elastic
scattering amplitude is indeed superfluous. 
 The differences for  $A_y$ and $iT_{11}$ can be very probably traced back
to the well known large sensitivity of these
observables to the $^3P_j$ components of the NN
interaction~\cite{physrep96} and different dynamics used by
us and in ~\cite{delt2005el}.

In Fig.~\ref{fig2} we show also by dotted blue lines results with the last
term in (\ref{eq11}) included. It is evident  that the  term
$-\left\langle {\vec p\vec q~} \right|\sum\limits_{\alpha '} {\int
{\left| {\alpha '} \right\rangle \left\langle
{\alpha '} \right|} } t_c^R PG_0S
 \sum\limits_{\alpha
''} {\int {\left| {\alpha ''}
\right\rangle \left\langle {\alpha ''} \right|} } T\left|
            {\Phi } \right\rangle$            
  is significant at low energies and that it deteriorates
  good description of data obtained with the first three terms.
  In ~\cite{wita_preprint} it was shown that at energies
  above $\approx 10$~MeV the contribution of that term to elastic scattering
  observables is negligible and  at $10$~MeV it starts to influence
  some spin observables. It is thus unavoidable below the breakup threshold
  to investigate how significant are effects of inclusion of
  the  fourth term 
 $ \left\langle
{\vec p\vec q~} \right|t_c^{R3d} PG_0 \sum\limits_{\alpha '} {\int
{\left| {\alpha '} \right\rangle \left\langle
  {\alpha '} \right|} } T\left| {\Phi } \right\rangle$
 in the elastic scattering transition amplitude. 
 Since the fifth term has a negative sign and contains partial wave
 contributions to the Coulomb
  t-matrix whose full 3-dimensional form is contained in the fourth term,  
  one would expect that they would at least 
  partially cancel each other and the inclusion of the fourth term should 
  restore at least partly the good description of data.

  The computation of  the fourth term with the 3-dimensional
  Coulomb t-matrix $t_c^{R3d}$, 
 can be done according to
expressions (D.9), (D.6), and (D.8) of Ref.~\cite{elascoul}.
It requires integrations over components of two vectors: 
over vector $\vec q$  in  (D.9), 
and over $\vec p~'$ or $\vec q_4$ in (D.6) or (D.8), respectively.
  Below the breakup threshold only channels
  $\alpha \ne \alpha_d$ contribute to (D.6).
  Since below the breakup threshold 
 the decomposition  (D.7) is superfluous, (D.8)
 provides the full contribution from $\alpha_d$ channels, obtained by replacing
 the second part of splitting  (D.7)  with the  left side of (D.7).
 The contributions from (D.6) and (D.8) must be determined numerically and
 this is the most time consuming part of the calculations.

 In Fig.~\ref{fig2} the indigo crosses show the results
obtained with all the terms in (\ref{eq11}) included. As expected 
 the fourth and fifth terms cancel each other to a large extent
 and a good description of data for the cross section and
 tensor analyzing powers is essentially regained. 

\begin{figure}
  \includegraphics[scale=0.7]{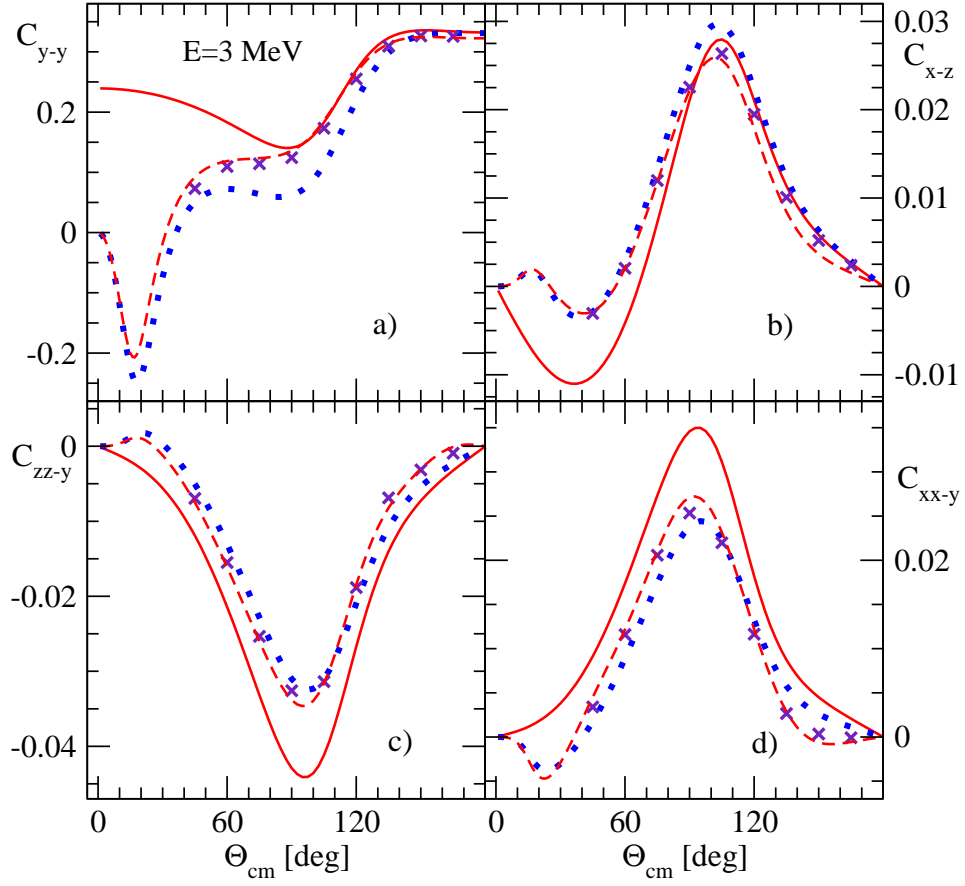}
  \caption{(color online) The same as in Fig.~\ref{fig2} but for selected
    spin correlation coefficients. 
  For description of lines see Fig.~\ref{fig2}.
 }
 \label{fig3}
\end{figure}

\begin{figure}
  \includegraphics[scale=0.7]{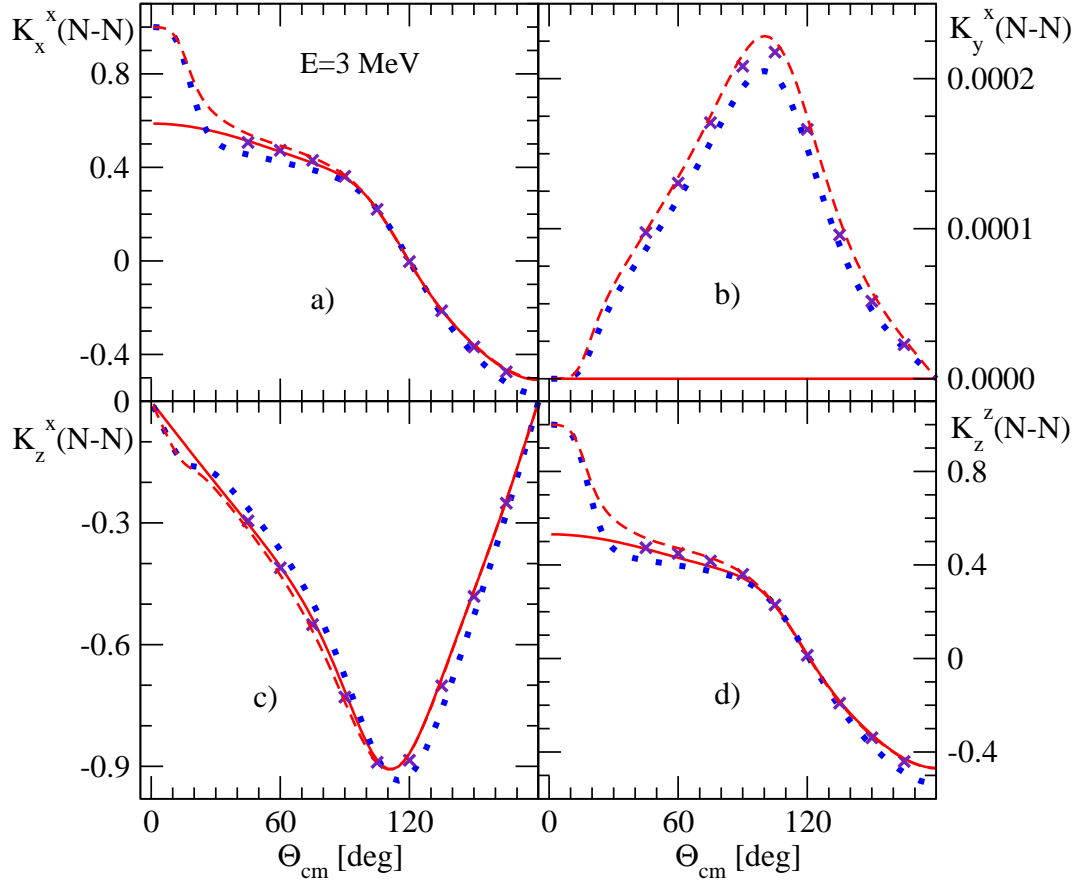}
  \caption{(color online) The same as in Fig.~\ref{fig2} but for selected
    proton to proton  spin transfer coefficients. 
  For description of lines see Fig.~\ref{fig2}.
 }
 \label{fig4}
\end{figure}

\begin{figure}
  \includegraphics[scale=0.7]{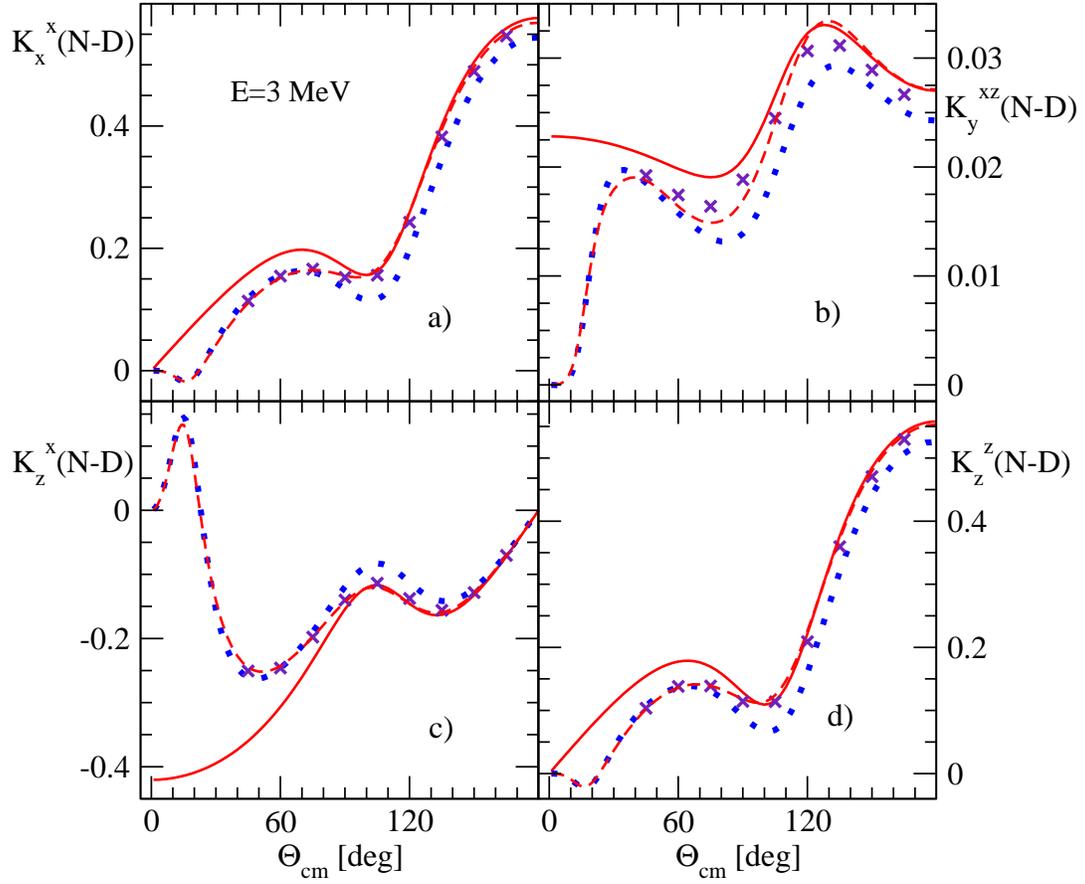}
  \caption{(color online) The same as in Fig.~\ref{fig2} but for selected
    proton to deuteron spin transfer coefficients. 
  For description of lines see Fig.~\ref{fig2}.
 }
 \label{fig5}
\end{figure}

\begin{figure}
  \includegraphics[scale=0.7]{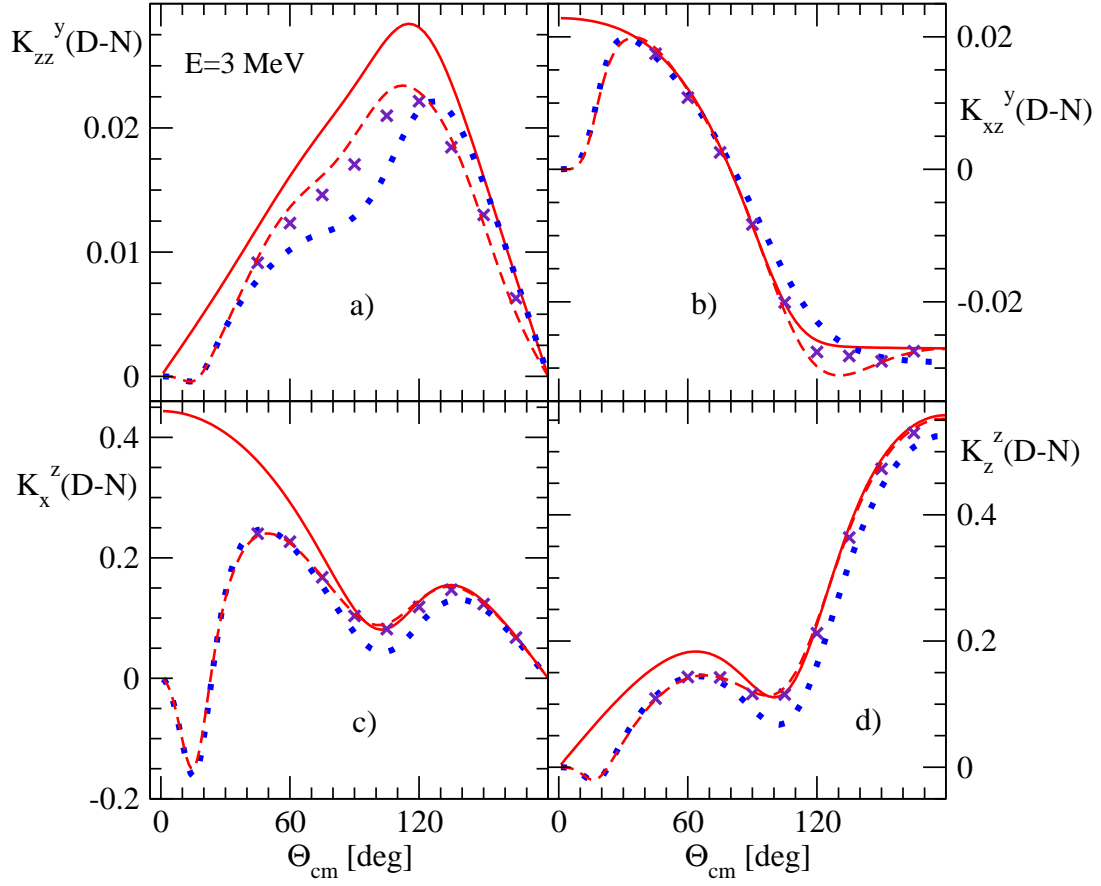}
  \caption{(color online) The same as in Fig.~\ref{fig2} but for selected
    deuteron to proton  spin transfer coefficients. 
  For description of lines see Fig.~\ref{fig2}.
 }
 \label{fig6}
\end{figure}

 To get an idea about the magnitude of the Coulomb force effects for other
 elastic scattering 
 observables we show in Figs.~\ref{fig3}-\ref{fig6} analogous
 predictions as in Fig.~\ref{fig2} but for selected spin correlations
 (Fig.~\ref{fig3}), proton to proton  (Fig.~\ref{fig4}),
 proton to deuteron (Fig.~\ref{fig5}), and deuteron to proton
 (Fig.~\ref{fig6}) spin transfer coefficients. The figures reveal
  a wide spectrum
 of importance and magnitude of the Coulomb force effects, dependent 
  on the observable.  
  For most of observables the effects are large in a wide range of angles, 
 for example for spin correlations from  Fig.~\ref{fig3} 
 and some of spin transfers ($K_y^x(N-N)$, $K_y^{xz}(N-D)$, $K_{zz}^{y}(D-N)$).
 For some large effects are restricted to forward region of angles below
 $\approx 90^o$ ($K_x^x(N-N)$, $K_z^{x}(N-N)$, $K_{z}^{z}(N-N)$,
 $K_x^x(N-D)$, $K_z^{x}(N-D)$, $K_{z}^{z}(N-D)$, $K_x^{z}(D-N)$, $K_{z}^{z}(D-N)$). 
 There are some interesting cases of observables which for the neutron-deuteron
 scattering vanish and become nonzero for the proton-deuteron interaction,
 as for example the nucleon to nucleon spin transfer coefficient 
 $K_y^x(N-N)$ shown in
 Fig.~\ref{fig4}. These nonzero values are due to a large charge
 independence breaking of pp and neutron-proton (np) interactions in isospin $t=1$ states,
 caused by the Coulomb  pp force.
 In our calculations we used  the charge dependent
 AV18 potentials, taking np and pp NN interactions of this model
 for the pd and nd systems.  In all isospin $t=1$
 states both total isospins of the 3N system $T=\frac {1} {2}$ and
 $T=\frac {3} {2}$ were taken into account.
  Vanishing of the $K_y^x(n-n)$ for nd scattering shows that
 the difference between np and pp NN AV18 potentials is too weak to induce
 nonzero values for this observable.

 The very interesting and most important effect seen  in all figures
 is that practically in all
 cases (large) effects caused by adding the fifth term to the elastic scattering
 transition amplitude are removed when including simultaneously the fourth term.
 In consequence, it is needless to account for these terms in
 elastic scattering amplitude what drastically simplifies and accelerates  
 determination of the Coulomb force effects.

Summarizing, we have shown that the two discussed approaches which enable to
include the long range Coulomb force in momentum-space pd 
 scattering calculations by applying a
 screening method have to provide the same results for all observables.
 In each method the cancellation between terms
containing 3-dimensional and partial wave decomposed Coulomb t-matrices is
decisive for establishing workable equations, whose structure is
 identical to the 
commonly used equations for neutron-deuteron scattering.
Solutions of these equations
together with four additional terms, two of which contain the 3-dimensional
Coulomb t-matrices, permit one 
to get the elastic scattering (and breakup) transition
amplitudes. In the approach based on the AGS equation 
it is unavoidable to perform
renormalization of the elastic scattering amplitudes before calculating 
observables. In the approach based
on the Faddeev equation such renormalization can be completely avoided.
 We have shown
numerically that the cancellation of last two terms 
in elastic scattering transition amplitude enables one to determine 
the pp Coulomb force effects
in the pd scattering nearly as easily as to compute 
  observables in neutron-deuteron scattering.

\clearpage

\acknowledgements
This research was supported in part by the Excellence
Initiative – Research University Program at the Jagiellonian
University in Krak\'ow. 
The numerical calculations were partly performed on the supercomputers of
the JSC, J\"ulich, Germany.

\begin{thebibliography}{99}

\bibitem{wita_preprint} H. Wita{\l}a, J. Golak, and R. Skibi\'nski,
  arXiv:2310.03433 [nucl.th].

\bibitem{delt_preprint} A. Deltuva, arXiv:2311.14605v1 [nucl.th].  

\bibitem{delt2005el} A. Deltuva, A. C. Fonseca, and P. U. Sauer,
  Phys. Rev. C{\bf{71}}, 054005 (2005).

\bibitem{delt2005br} A. Deltuva, A. C. Fonseca, and P. U. Sauer,
  Phys. Rev.  C{\bf{72}}, 054004 (2005).  

\bibitem{elascoul} H. Wita{\l}a, R. Skibi\'nski, J. Golak,
  W. Gl\"ockle,
  Eur. Phys. Journal A{\bf{41}}, 369 (2009).

\bibitem{brcoul} H. Wita{\l}a, R. Skibi\'nski, J. Golak,
  W. Gl\"ockle,
  Eur. Phys. Journal A{\bf{41}}, 385 (2009).  
  
\bibitem{AGS} E. O. Alt, P. Grassberger, W. Sandhas, Nucl. Phys. {\bf B2},
  167 (1967).

\bibitem{gloeckle83} W. Gl\"ockle, The Quantum Mechanical Few-Body Problem,
  Springer Verlag 1983.

\bibitem{physrep96}  W. Gl\"ockle, H. Wita{\l}a, D. H\"uber, H. Kamada, J.
 Golak, Phys. Rep. {\bf{274}}, 107 (1996).

\bibitem{skib2009}  R. Skibi\'nski, J. Golak, H. Wita{\l}a,  and W.Gl\"ockle, 
  Eur. Phys. Journal A{\bf{40}}, 215  (2009).

\bibitem{Alt78} E. O. Alt, W. Sandhas, and H. Ziegelmann, Phys. Rev. {\bf C
17}, 1981 (1978).

\bibitem{chen72} J.C.Y. Chen and A.C. Chen,
in Advances of Atomic and Molecular Physics,
 edited by D. R. Bates and J. Estermann ( Academic, New York, 1972), Vol. 8.

\bibitem{kok1980} L. P. Kok, H. van Haeringen,  P
  hys. Rev. C{\bf{21}}, 512 (1980).

\bibitem{ford1964} W.F. Ford, Phys. Rev. {\bf 133}, B1616 (1964).

\bibitem{ford1966} W.F. Ford, J. Math. Phys. {\bf 7}, 626 (1966).

\bibitem{taylor1} J.R. Taylor,  Nuovo Cimento {\bf B23}, 313 (1974).

\bibitem{taylor2} M.D. Semon and J.R. Taylor,
  Nuovo Cimento {\bf A26}, 48 (1975).

\bibitem{kok1981} L. P. Kok, H. van Haeringen,
  Phys. Rev. Lett. {\bf{46}}, 1257 (1981).

\bibitem{haer1985} H. van Haeringen, Charged Particle Interactions,
  Theory and Formulas, (Coulomb Press, Leyden, 1985).

\bibitem{av18}  R. B. Wiringa, V. G. J. Stoks, R. Schiavilla, 
  Phys. Rev. C{\bf 51}, 38 (1995).

\bibitem{shimizu1995} S. Shimizu {\em et al}.,  Phys. Rev. C {\bf 52}, 
  1193 (1995).

\end{thebibliography}
\end{document}